\documentclass[a4paper]{article}

\usepackage{INTERSPEECH2020}
\usepackage{hyperref}

\title{Towards a Competitive End-to-End Speech Recognition \protect\\ for CHiME-6 Dinner Party Transcription}
\name{Andrei Andrusenko$^{1,*}$\thanks{$^*$Equal contribution.}, Aleksandr Laptev$^{1,*}$, Ivan Medennikov$^{1,2}$}
\address{
  $^1$ITMO University, St. Petersburg, Russia\\
  $^2$STC-innovations Ltd, St. Petersburg, Russia}
\email{\{andrusenko, laptev, medennikov\}@speechpro.com}

\begin{document}

\maketitle
\begin{abstract}
  While end-to-end ASR systems have proven competitive with the conventional hybrid approach, they are prone to accuracy degradation when it comes to noisy and low-resource conditions. 
  In this paper, we argue that, even in such difficult cases, some end-to-end approaches show performance close to the hybrid baseline. To demonstrate this, we use the CHiME-6 Challenge data as an example of challenging environments and noisy conditions of everyday speech. We experimentally compare and analyze CTC-Attention versus RNN-Transducer approaches along with RNN versus Transformer architectures. We also provide a comparison of acoustic features and speech enhancements.
  Besides, we evaluate the effectiveness of neural network language models for hypothesis re-scoring in low-resource conditions.
  Our best end-to-end model based on RNN-Transducer, together with improved beam search, reaches quality by only 3.8\% WER abs. worse than the LF-MMI TDNN-F CHiME-6 Challenge baseline. With the Guided Source Separation based training data augmentation, this approach outperforms the hybrid baseline system by 2.7\% WER abs. and the end-to-end system best known before by 25.7\% WER abs.
\end{abstract}

\section{Introduction}

In recent years, the active development of deep learning techniques has allowed researchers to make a great performance improvement of Automatic Speech Recognition (ASR) systems. Although conventional hybrid ASR systems \cite{hinton_hmm-dnn} have been remaining preferred for a long time, their ever-increasing difficulty of construction and training \cite{medennikov_stc_2019} has led to an interest in end-to-end approaches \cite{graves_connectionist_2006,chan2015listen}. Unlike HMM-DNN-based, these methods are trained to directly map an input acoustic feature sequence to a sequence of characters with a single objective function that is highly relevant to the final evaluation criteria, in particular, Word Error Rate (WER). The freedom from the necessity of intermediate modeling, such as acoustic and language models with pronunciation lexicons, makes the process of building the system clearer and more straightforward.

Also, there is an informal competition in accuracy between these systems \cite{L_scher_2019}. In most speech recognition tasks with good sound conditions and speech quality (e.g. LibriSpeech \cite{panayotov_librispeech:_2015}), end-to-end models' performance is great \cite{synnaeve2019endtoend,li2019jasper,Zeyer_2018}. However, in case of robust far-field speech recognition in noisy environments and with low resources, such models face problems \cite{bataev_2018}. There is a series of CHiME Challenges, which task is to encourage researchers to solve the problem of speech recognition in such conditions. According to the previous CHiME-5 Challenge \cite{barker2018fifth} reports, the leading positions in quality can be reached only by conventional hybrid systems \cite{chime5_1,chime5_2,chime5_3}. Attempts to get a somehow comparable result for end-to-end systems have not shown any notable success \cite{chime5_e2e}.

There is a new line of research aimed at solving a similar problem of the multichannel robust speech recognition task using the end-to-end approaches. Works such as \cite{yalta2018cnnbased,subramanian2019investigation,Chang2019MIMOSpeechEM} demonstrate the effectiveness of joint training of neural-network-based front-end (speech enhancement) and back-end (speech recognition) models compared to the use of separate models. However, if the utterance boundaries are given, currently, the most effective approach is the speech enhancement technique based on spatial GMM blind source separation, named Guided Source Separation (GSS) \cite{boeddecker_front-end_2018}. Thus, while maintaining the flexibility of end-to-end modeling, combining the same GSS front-end and end-to-end models might be capable of achieving the quality of conventional hybrid systems for a difficult CHiME-6 dinner party recognition task.

This paper describes an investigation of the aforementioned conjunction of techniques. We used the sixth CHiME Challenge \footnote{\url{https://chimechallenge.github.io/chime6/}} data, as it has an additional accurate array synchronization, and the baselines for speech enhancement and recognition. Our experimental setup was focused on the replacement of the hybrid recognition system from the baseline with an end-to-end system. We explored the effectiveness of RNN and Transformer architectures along with CTC-Attention and RNN-Transducer (RNN-T) training-decoding approaches.

This study also considers some extensions for the basic RNN-T model:
\begin{itemize}
    \item Transformer-transducer \cite{Tian_2019,yeh2019transformertransducer}. The idea behind this approach is to replace conventional LSTM/BLSTM encoder and predictor modules of the basic RNN-T implementation with the Transformer networks, which have proven effective in sequence modeling tasks.
    \item Improved beam search for RNN-T decoding \cite{jain2019rnnt}. This is a modification of the standard beam search algorithm aimed at improving the computational decoding efficiency by using several hypothesis pruning heuristics.
    \item ASGD Weight-Dropped LSTM language model (AWD-LSTM-LM) \cite{merity2017regularizing} for hypotheses re-scoring. This approach is to apply various regularization strategies to address overfitting caused by a low text data amount.
    \item The use of GSS-based speech enhancement for training and testing data \cite{zorila_investigation_2019,Medennikov_chime6}.
\end{itemize}

Our main findings are available as an ESPnet CHiME-6 recipe\footnote{\url{https://github.com/espnet/espnet/tree/master/egs/chime6/asr1}}.

The rest of the paper is organized as follows. Section~\ref{section:e2e} provides an overview of common end-to-end ASR approaches. The CHiME-6 data and baseline are described in Section~\ref{section:chime6}. Section~\ref{section:exp} provides an experimental setup for the data and the strategies along with a discussion of our findings. Finally, Section~\ref{section:conclusion} presents our conclusions and future work.

\section{End-to-end ASR overview}
\label{section:e2e}

\subsection{CTC-Attention}

Connectionist temporal classification (CTC), originally introduced in \cite{graves_connectionist_2006}, in the ASR field is implied as a type of neural network output and an associated scoring function for training sequence-to-sequence (Seq2Seq) models. This approach is intended to relax the requirement of one-to-one mapping (alignment) between the input and output sequences. It uses an intermediate label representation with a special ``blank'' symbol indicating that no label is seen. It is also used when the label is repeated. One of the key features worth mentioning is the conditional label independence at the inference process, which hinders the effective usage of CTC models without an external language model.

The attention-based encoder-decoder approach, in contrast to CTC, incorporates contextual information by using both the input frames and the history of the target label for the inference process \cite{chan2015listen}. It learns to predict the alignment between frames and labels using the attention mechanism. While generally performing better than CTC, attention-based approaches are more susceptible to noise and require more effort to train. It is proven effective to combine these approaches to alleviate their shortcomings and improve recognition quality \cite{kim_joint_2017}.

Typical CTC-Attention architecture contains a deep RNN (mostly, bi- or unidirectional LSTM) encoder and a unidirectional RNN decoder with the optional use of an external language model (LM).

\subsection{Transformer}

Examining ASR Seq2Seq models, it is worth mentioning the Transformer architecture. Originally proposed for neural machine translation (NMT) \cite{attention_2017}, it delivers generally better accuracy compared to RNN \cite{karita_comparative_2019}. The Transformer incorporates self-attention to utilize sequential information in contrast to the recurrent connection employed in RNN. Similar to traditional CTC-Attention, it uses joint CTC- and attention-based decoding.

The transformer architecture consists of both deep Multi-head attention (MHA) encoder and decoder. To represent the time location, linear or convolutional positional encoding is applied before the encoder and decoder modules.

\subsection{Neural Transducer}

\begin{figure}[ht]
  \centering
  \includegraphics[scale=0.17]{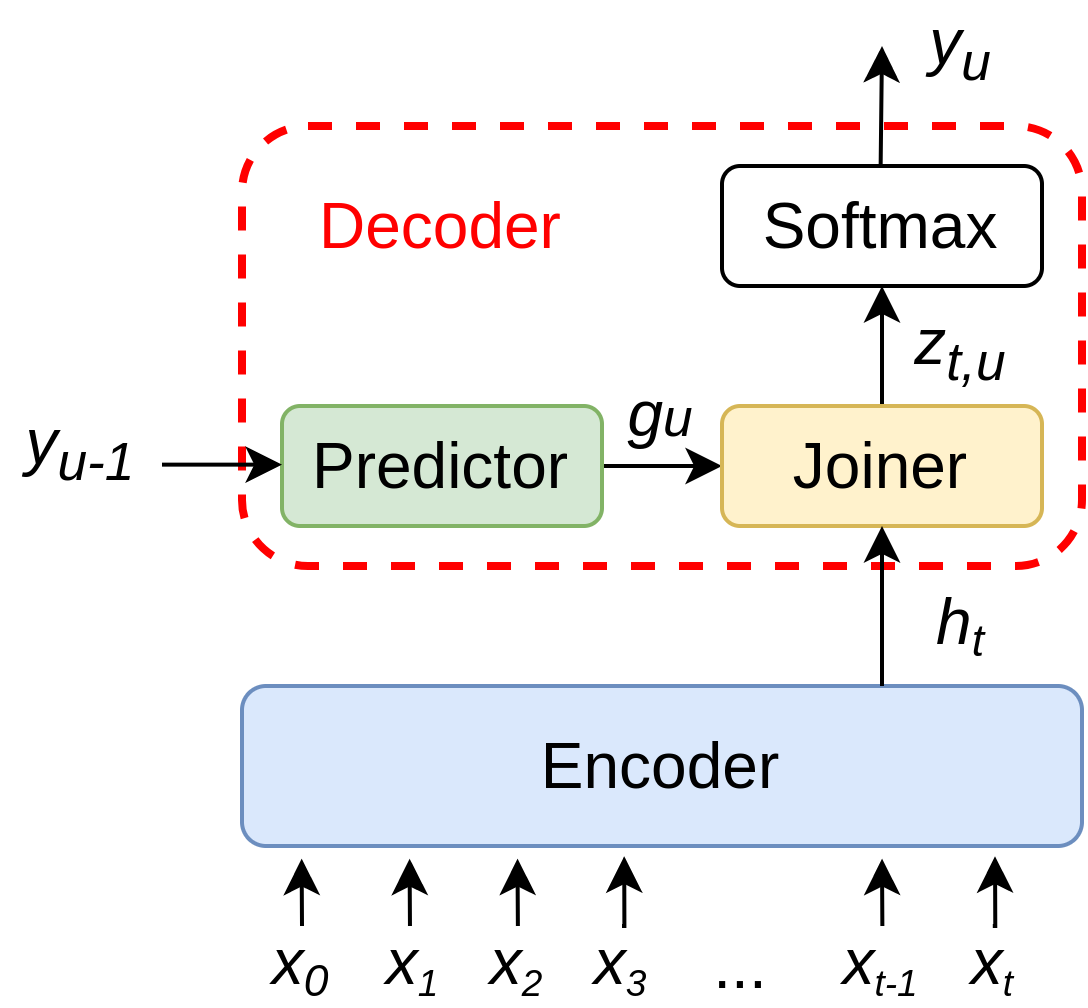}
  \caption{Neural Transducer.}
  \label{fig:rnnt}
\end{figure}

The main alternative to the CTC-Attention approach is the neural transducer \cite{Graves_transducer}. In many details, this approach is similar to the CTC. They both use the ``blank'' label and compute the probability of all possible alignment paths during training to obtain the probability of the entire target transcription. Architecturally, having essentially the same encoder, the decoder of the neural transducer is different from the attention-based. It consists of:
\begin{itemize}
    \item A prediction network, which plays the role of a language model.
    \item A joint network, intended for aligning audio and label sequences.
\end{itemize}

The standard neural transducer structure is presented in Figure \ref{fig:rnnt}. An input acoustic feature sequence \(\mathbf{x}=(x_0,x_1,x_2,...,x_t)\) is passed to the Encoder and converted to a sequence of embeddings \(\mathbf{h}=(h_0,h_1,h_2,...,h_{T^{'}})\). In most cases, the Encoder performs subsampling by using convolutional layers, therefore \(T\leq{T^{'}}\). Next, the Joiner, which is a shallow fully connected (FC) neural network, receives the current Encoder embedding $h_t$ and predictor embedding $g_u$ to yield logits $z_{t,u}$. The most probable label $y_u$ is determined by the Softmax. It is important that, since the Predictor practically works like a neural LM, it accepts only a non-blank $y_{u-1}$. For the blank $y_{u-1}$, it yields $g_u$ as if for the last non-blank $y_{u-1}$.

Traditionally, most neural transducer studies investigate RNN-transducer. It consists of deep bi- or unidirectional LSTM encoder and a unidirectional LSTM predictor. According to \cite{Tian_2019}, the neural transducer is able to reach better accuracy with the use of the Transformer as an architecture of encoder or predictor. Also, the Transformer-Transducer from \cite{yeh2019transformertransducer} outperformed the basic RNN-T model for the Librispeech task.

Apart from architecture improvement, the neural transducer beam search is also can be improved. Work \cite{jain2019rnnt} introduces \textit{expand\_beam} and \textit{state\_beam} parameters to explicitly limit the beam size in the decoding process, and thus, to increase decoding efficiency. These parameters allow the search to reject ``bad'' hypotheses and to choose only the most probable, in terms of confidence, in the predictions of the neural network.

\section{CHiME-6 dinner party transcription}
\label{section:chime6}

The previous CHiME-5 Challenge featured fully transcribed speech materials collected with multiple 4-channel microphone arrays from real dinner parties that have taken place in real homes. Conversational speech with a large amount of overlapped segments recorded in reverberant and noisy conditions significantly complicates recognition. Details on the Challenge can be found in \cite{barker2018fifth}.

CHiME-6 Challenge \cite{chime6} used the same recordings as the previous one, but improved initial data preparation and established the following strong baselines:

\begin{itemize}
    \item Two stages array synchronization by \textit{frame-dropping} and \textit{clock-drift}. This allowed to obtain utterances with the same start and end time on every device.  
    \item GSS-based speech enhancement \cite{boeddecker_front-end_2018} applied to multiple arrays.
    \item Factorized time-delayed neural network (TDNN-F) acoustic model, trained with lattice-free maximum mutual information (LF-MMI), from Kaldi\footnote{\url{https://github.com/kaldi-asr/kaldi}} toolkit.
\end{itemize}

This baseline demonstrated 51.76\% WER on the development set by using \textit{train\_worn\_simu\_u400k\_cleaned\_rvb} training set, which was obtained by various augmentations of 40 unique training data. Specific rooms simulation, speed, and volume perturbation increased the total amount of data to about 1400 hours.

\section{Experimental Setup}
\label{section:exp}

We used the ESPnet\footnote{\url{https://github.com/espnet/espnet}} speech recognition toolkit \cite{watanabe_espnet:_2018} as the main framework for our experiments. It supports most of the basic end-to-end models and training pipelines.

\subsection{End-to-end modeling}

The first step of our investigation was to discover the most suitable end-to-end architecture for solving the task under consideration. Approaches like joint CTC-Attention, RNN-Transducer, and Transformer are the most popular for ASR tasks. Quasi-optimal configurations of end-to-end architectures for our task are presented in Table~\ref{tab:models_configuration}. ``T-T'' and ``dp'' stand for Transformer-Transducer and dropout, respectively. The Transformer network is used only in the Encoder in this architecture.

\begin{table}[ht]
  \caption{Models configuration}
  \label{tab:models_configuration}
  \centering
  \begin{tabular}{c c}
    \toprule
    \textbf{CTC-Attention} \\
    Encoder &  VGG-BLSTM, 6-layer 512-units, dp 0.4\\
    Attention & 1-head 256-units, dp 0.4 \\
    Decoder &  LSTM, 2-layer 256 units, dp 0.4\\
    \midrule
    \textbf{RNN-T} \\
    Encoder &  VGG-BLSTM, 6-layer 512-units, dp 0.4\\
    Predictor & LSTM, 2-layer 256-units, dp 0.4 \\
    Joiner &  FC 256 units\\
    \midrule
    \textbf{T-T} \\
    Encoder &  MHA, 12-layer 1024-units, dp 0.4\\
    Attention & 8-head 512-units, dp 0.4 \\
    Predictor & LSTM, 2-layer 256-units, dp 0.4 \\
    Joiner &  FC 256 units\\
    \midrule
    \textbf{Transformer} \\
    Encoder &  MHA, 12-layer 1024-units, dp 0.5\\
    Attention & 4-head 256-units, dp 0.5 \\
    Decoder &  MHA, 2-layer 1024 units, dp 0.5 \\
    \bottomrule
  \end{tabular}
\end{table}

For all the presented approaches, we used a convolutional network (CNN) in front of the encoder. It consists of four Visual Geometry Group (VGG) CNN layers designed to compress the input frames in time scale by the factor of 4. Thus, after applying this network, the output features represent the transformed information from 4 initial frames. Although not originally intended, this approach allowed the model to converge more stably.

\subsection{Experimental evaluation}

We used the official data from the Kaldi baseline recipe: \textit{train\_worn\_simu\_u400k\_cleaned\_rvb} and \textit{dev\_gss\_multiarray} (or \textit{dev\_gss12} in our simplified notation) as the training and development sets, respectively.

Characters (26 letters of the English alphabet and 7 auxiliary symbols) were used as acoustic units. Other versions of word units, namely position-depending letters, Byte Pair Encoding (BPE) \cite{sennrich2015neural} with different numbers of units 500, 1000, 2000 led only to performance degradation. This might be due to the lack of training data for such large unit numbers.

The next step was to choose the input acoustic features. The hybrid baseline model used 40-dimensional high-resolution MFCC vectors (hires) concatenated with 100-dimensional i-vectors. However, such features may not be the best option for an end-to-end model. We considered the following features: 80-dimensional log-Mel filterbank coefficients (fbank) with or without 3-dimensional pitch features and 512-dimensional wav2vec context network output vectors \cite{schneider_wav2vec:_2019}. A comparison of these acoustic features with the RNN-T model configured as in Table~\ref{tab:models_configuration} is shown in Table~\ref{tab:feats}. All decoding results were obtained using the beam size of 10.

\begin{table}[ht]
  \caption{RNN-T features comparison}
  \label{tab:feats}
  \centering
  \begin{tabular}{c c c}
    \toprule
    \textbf{Feats} & \textbf{Dimension} & \textbf{WER(\%)} \\
    \midrule
    wav2vec & 512 & 68.3 \\
    \midrule
    hires+i-vectors (baseline)  & 40+100 & 64.1 \\
    \midrule
    hires & 40 & 63.6 \\
    \midrule
    fbank & 80 & 60.4 \\
    \midrule
    fbank+pitch & 80+3 & 59.7 \\
    \bottomrule
  \end{tabular}
\end{table}

Limited computing power did not allow us to train the wav2vec features extractor for a sufficient number of epochs (the model was trained only in 6 epochs), which apparently caused such a bad result. The underperforming of hires+i-vectors against the single hires might be due to the VGG network usage. All further experiments were carried out using fbank+pitch features.

Having settled on 33 character acoustics units and the fbank+pitch acoustics features, we compared the end-to-end architectures from the Table~\ref{tab:models_configuration}. We also used SpecAugment \cite{Park_2019} to further augment the training data. The results of this comparison are presented in Table~\ref{tab:models}.

\begin{table}[ht]
  \caption{Comparison of end-to-end models}
  \label{tab:models}
  \centering
  \begin{tabular}{c c c c c}
    \toprule
    \ & \textbf{CTC-Att} & \textbf{RNN-T} & \textbf{T-T} & \textbf{Transf.}  \\
    \midrule
    \textbf{WER(\%)} & 73.8 & \textbf{55.5} & 60.3 & 86.8 \\
    \bottomrule
  \end{tabular}
\end{table}

The results show that the RNN-T demonstrates the best WER in this task. It is worth noting the great positive impact of SpecAugment. The applying of this augmentation reduced WER by 4.2\% for the RNN-T model. This suggests the importance of augmentation methods for low-resource tasks. Models that utilize the Transformer architecture (T-T, Transformer) perform worse than LSTM due to severe overfitting.

We also faced a problem in the re-scoring process using external RNN-based language models. Character and word-based LSTM LMs (1-layer 1024-units) were trained on the training data transcriptions only. However, the use of these LMs for hypothesis re-scoring in the beam search algorithm for RNN-T only resulted in reduced recognition accuracy. We also used the pre-trained AWD-LSTM-LM, which showed a significant WER improvement for lattice re-scoring in the STC CHiME-6 system \cite{Medennikov_chime6}. Unfortunately, none of these models improved accuracy. Apart from the beam search rescoring, we tried to use the n-best rescoring approach, but the n-best hypotheses produced from the development set utterances did not have significant variability in words. Thus, this rescoring method did not improve the result either. We have two assumptions for this to happen. The first one is that in case of a low-resource task, with properly tuned Predictor network parameters can perform sufficiently good without using any re-scoring techniques. The second one is that the beam search algorithm we used is not optimal for the hypothesis re-scoring. For example, in the case of a hybrid system decoding, there are separate acoustic and language model scores of a word unit. And in the corresponding re-scoring algorithm, the language score is re-weighted according to the external LSTM LM. But in the case of RNN-T, there is only one single score for word unit. The default ESPnet re-scoring is simply a weighted sum of the external LSTM LM and RNN-T scores. The results of hypothesis re-scoring with an external NNLM are demonstrated in Table~\ref{tab:external_nnlm}. We used the beam size of 10 and the NNLM weight of 0.3 for all experiments.

\begin{table}[ht]
  \caption{External NNLM for RNN-T beam search}
  \label{tab:external_nnlm}
  \centering
  \begin{tabular}{c c}
    \toprule
    \textbf{NNLM} & \textbf{WER(\%)} \\
    \midrule
    - & 55.5 \\
    \midrule
    Char NNLM & 57.5 \\
    \midrule
    Word NNLM & 57.4 \\
    \midrule
    AWD-LSTM & 56.3 \\
    \bottomrule
  \end{tabular}
\end{table}

The improved beam search algorithm, implemented according to \cite{jain2019rnnt}, demonstrated results beyond expectations. Values \textit{expand\_beam} and \textit{state\_beam} handpicked as 2 and 1, respectively, allowed to accelerate the decoding process by an average of 15-20\% with a simultaneous improvement in the recognition quality by 0.5 absolute WER, thus delivering 55.00 WER for the \textit{dev\_gss12}. We assume that the WER improvement was due to a decrease in the impact of the Predictor overfitting.

\begin{table}[ht]
  \caption{The final models' performance comparison on the development set.}
  \label{tab:final_all}
  \centering
  \begin{tabular}{c c}
    \toprule
    \textbf{Model} & \textbf{WER(\%)} \\
    \midrule
    Joint CTC/Attention E2E \cite{chime5_e2e} & 82.1 \\
    CNN-based Multichannel E2E \cite{yalta2018cnnbased} & 80.7 \\
    CHiME-6 TDNN-F baseline \cite{chime6} & 51.7 \\
    \midrule
    RNN-T + \textit{dev\_gss12} & 55.0 \\
    RNN-T + \textit{train\_gss} + \textit{dev\_gss12} & 52.6 \\    
    RNN-T + \textit{train\_gss} + \textit{dev\_gss24} & 49.0 \\
    \midrule
    Hybrid system (n-gram LM) \cite{Medennikov_chime6} & 36.8 \\
    Hybrid system (AWD-LSTM-LM) \cite{Medennikov_chime6} & 33.8 \\
    \bottomrule
  \end{tabular}
\end{table}

We also applied additional GSS-based speech enhancement for the training (\textit{train\_gss}) data according to \cite{Medennikov_chime6} as well as improved 24-microphone GSS enhancement for the development data (\textit{dev\_gss24}).

The final comparison of some of the currently published end-to-end models, the official baseline, and our RNN-T models is presented in Table~\ref{tab:final_all} (only the CHiME-6 development data WER results were considered). Note that results from \cite{chime5_e2e} and \cite{yalta2018cnnbased} are obtained for the CHiME-5 data, i.e. without improvements mentioned in Section~\ref{section:chime6}. To demonstrate a gap between our system and one of the best hybrid systems known to us at the moment, we also placed in the comparison the best single model from the STC CHiME-6 system \cite{Medennikov_chime6}.

\section{Conclusion}
\label{section:conclusion}

In this work, we presented an end-to-end systems exploration of the robust far-field speech recognition in noisy environments and low resources such as the CHiME-6 dinner party transcription task. We showed that the combination of powerful GSS-based speech enhancement and the use of the RNN-Transducer model is able to achieve competitive results with hybrid systems. With the improved beam search algorithm and addition speech enhancement, our system outperformed the hybrid baseline system by 2.7\% WER abs.

Further research might be the study on how, for the task under consideration, to achieve a recognition accuracy improvement when re-scoring hypotheses using external NNLM. Also, it would be interesting to incorporate GSS-based enhancement in the end-to-end pipeline to train the system jointly.

\section{Acknowledgements}
We would like to show our sincere gratitude to A. Mitrofanov, I. Podluzhny, and A. Romanenko for sharing their developments in external NNLM modeling for CHiME-6 data. We also want to thank the ESPnet development team for the excellent open-source ASR toolkit that helped us doing this research.

This work was partially financially supported by the Government of the Russian Federation (Grant 08-08).

\bibliographystyle{IEEEtran}

\bibliography{mybib}

\end{document}